\documentclass[aps, prl,10pt,twocolumn,superscriptaddress,nofootinbib]{revtex4-1}
\usepackage{mathrsfs, amssymb, amsmath}  
\usepackage{epsfig, cancel}
\usepackage{latexsym}
\usepackage{natbib, comment}
\usepackage{url}
\usepackage{dcolumn}
\usepackage{multirow}
\usepackage{color}
\usepackage{cancel}
\usepackage{soul}
\usepackage[normalem]{ulem}
\usepackage{amsfonts,amssymb,amsmath, txfonts}
\usepackage{graphicx,epsfig}
\usepackage{psfrag}
\usepackage{hyperref}
\hypersetup{colorlinks=true}
\usepackage{mathtools}
\usepackage{enumitem}
\usepackage{float}
\usepackage[dvipsnames]{xcolor}
\usepackage{xcolor}
\hypersetup{ linktoc=all,
    colorlinks, linkcolor={blue},
    citecolor={darkred}, urlcolor={darkgreen}
}
\definecolor{rosy}{RGB}{230,235,252}
\definecolor{myframetitle}{RGB}{90,89,170}
\definecolor{myblocktitle}{RGB}{140,185,249}
\definecolor{mytitle}{RGB}{10,80,26}

\definecolor{darkgreen}{RGB}{27,130,45}
\definecolor{darkblue}{rgb}{0,0,0.3}
\definecolor{darkred}{rgb}{0.7,0,0}

\definecolor{light gray}{RGB}{220,220,220}
\definecolor{dark purple}{RGB}{108,0,217}
\definecolor{pink}{RGB}{190,20,100}
\definecolor{orang}{RGB}{193,63,0}
\definecolor{green}{RGB}{11,98,17}
\definecolor{darkpink}{RGB}{153,0,76}
\definecolor{bluegreen}{RGB}{0,102,102}
\definecolor{greenlagan}{RGB}{0,102,0}
\definecolor{redgreen}{RGB}{102,102,0}
\definecolor{Redgreen}{RGB}{153,76,0}
\definecolor{vividviolet}{rgb}{0.62, 0.0, 1.0}
\definecolor{amaranth}{rgb}{0.9, 0.17, 0.31}
\definecolor{palatinateblue}{rgb}{0.15, 0.23, 0.89}
\definecolor{brightpink}{rgb}{1.0, 0.0, 0.5}
\definecolor{cornflowerblue}{rgb}{0.39, 0.58, 0.93}
\definecolor{deepcarminepink}{rgb}{0.94, 0.19, 0.22}
\definecolor{radicalred}{rgb}{1.0, 0.21, 0.37}

\def\red{\textcolor{red}}

\usepackage[english]{babel}

\usepackage[letterpaper,top=2cm,bottom=2cm,left=3cm,right=3cm,marginparwidth=1.75cm]{geometry}
\def\H0{{\text{H}\hspace*{-2.05mm}\text{H} 0\hspace*{-1.35mm}0\ }}

\def\be{\begin{equation}}
\def\ee{\end{equation}}
\def\beq{\begin{equation}}
\def\eeq{\end{equation}}
\def\bea{\begin{eqnarray}}
\def\eea{\end{eqnarray}}
\newcommand{\dd}{\textrm{d}}

\begin{document}

\title{How much has DESI dark energy evolved since DR1?}

\author{Eoin \'O Colg\'ain}
\email{eoin.ocolgain@atu.ie}
\affiliation{Atlantic Technological University, Ash Lane, Sligo, Ireland}
\author{Saeed Pourojaghi} 
\email{s.pouri90@gmail.com}
\affiliation{School of Physics, Institute for Research in Fundamental Sciences (IPM), P.O.Box 19395-5531, Tehran, Iran}
\author{M. M. Sheikh-Jabbari} 
\email{shahin.s.jabbari@gmail.com}
\affiliation{School of Physics, Institute for Research in Fundamental Sciences (IPM), P.O.Box 19395-5531, Tehran, Iran}
\author{Lu Yin}
\email{yinlu@shu.edu.cn}
\affiliation{Department of Physics, Shanghai University, Shanghai, 200444,  China}

\begin{abstract}

DESI has reported a dynamical dark energy (DE) signal based on the $w_0 w_a$CDM model that is in conflict with Hubble tension. Recalling that the combination of DESI DR1 BAO and DR1 full-shape (FS) modeling are consistent with $\Lambda$CDM, in this letter we comment on the status of fluctuations in DR1 BAO documented in \cite{DESI:2024mwx, Colgain:2024xqj} in the DR2 update. In particular, we note that neither DR1 BAO nor DR2 BAO nor DR2 BAO+CMB confronted to the $w_0 w_a$CDM model with relaxed model parameter priors confirm  late-time accelerated expansion today. Translating DESI BAO constraints into flat $\Lambda$CDM constraints, we observe that the LRG1 constraint remains the most prominent outlier, a distinction now held jointly with ELG1, LRG2 switches from smaller to larger $\Omega_m$ values relative to Planck-$\Lambda$CDM, and ELG data drive the relatively low $\Omega_m$ in the full DR2 BAO. We observe that one cannot restore $w_0 = -1$ within one $1 \sigma$ by removing either LRG1 or ELG1 or LRG2, but LRG2 in DR2, in contrast to LRG1 in DR1, now has the greatest bearing on $w_0 > -1$. We conclude that BAO has yet to stabilise, but the general trend is towards greater consistency with DESI DR1 FS modeling results, where there may be no dynamical DE signal in DESI data alone.

\end{abstract}

\maketitle

\section{Introduction}

Recent observations of a statistically significant dynamical dark energy (DE) \cite{DESI:2024mwx, DESI:2024hhd, DESI:2025zgx} (see \cite{DES:2024jxu, Rubin:2023ovl} for earlier claims) based on the CPL model \cite{Chevallier:2000qy, Linder:2002et} are problematic on a number of fronts. First, local (model-independent) $H_0$ determinations are biased to $H_0 > 70$ km/s/Mpc values (see \cite{Skara:2019usd, DiValentino:2021izs, Abdalla:2022yfr, DiValentino:2025sru} for reviews). To the extent of our knowledge, it has been appreciated since 2018, perhaps even earlier, that a DE equation of state in the traditional quintessence regime $w_{\textrm{DE}}(z) > -1$ exacerbates the Hubble tension  \cite{Vagnozzi:2018jhn, Vagnozzi:2019ezj, Alestas:2020mvb, Banerjee:2020xcn, Lee:2022cyh}. The result holds in simple $w_{\textrm{DE}}(z)$ parameterizations, e. g. $w$CDM \cite{Vagnozzi:2019ezj}, $w_0 w_a$CDM \cite{Vagnozzi:2018jhn, Alestas:2020mvb}, and more general field theories \cite{Banerjee:2020xcn, Lee:2022cyh}. See appendix for a simple analytic argument why this must be so.

In particular, it was observed in  \cite{Lee:2022cyh} that $w_0 := w_{\textrm{DE}}(z=0) > -1$ hinders a resolution to Hubble tension even if $w_{\textrm{DE}}(z) < -1$ at $z > 0$. One sees the problem clearly in DESI DR2 results \cite{DESI:2025zgx}: in the combination BAO+CMB+SNe, for different SNe samples, respectively  Pantheon+ \cite{Brout:2022vxf}, DES  \cite{DES:2024jxu} and Union3 \cite{Rubin:2023ovl},  as $w_0$ increases from $w_0 = -1$, i. e. the $\Lambda$CDM deviation becomes more statistically significant, $H_0$ decreases (we quantify this problem in the appendix). This observation pertains to any combination of datasets with $w_0 > -1$ and is more general than the CPL model, applying to DE models fitted to DESI data in lieu of CPL, e. g. \cite{DESI:2024kob, DESI:2025fii}. {Note, the $w_0-H_0$ anti-correlation in DESI results tells us that the DESI dynamical DE claim and a high local $H_0$, or $H_0$ tension, are contradictory. Observe also that this is a stronger statement than  late-time dynamical DE cannot resolve $H_0$ tension \cite{Bernal:2016gxb,Lemos:2018smw,  Krishnan:2021dyb, Cai:2021weh, Keeley:2022ojz, Gomez-Valent:2023uof}.}  

The second problem is that given the inevitable unknown unknowns in cosmology, one can only trust a result if one sees it \textit{independently} in distinct datasets. The prototypical example of this is late-time accelerated expansion, which necessitates the presence of DE modeled through $\Lambda$ in the $\Lambda$CDM model, and is seen independently across virtually all observables; not seeing support for $\Lambda$ undermines the observable. In contrast, the DESI dynamical DE claim \cite{DESI:2024mwx, DESI:2024hhd, DESI:2025zgx} is only statistically significant when datasets are combined; the significance depends on the data combinations  \cite{Giare:2025pzu} and is especially sensitive to SNe \cite{Capozziello:2025qmh}. In defense of combining datasets, it is well recognized that BAO, SNe and CMB only weakly constrain the CPL model on their own. This is in part due to the fact that CPL can be (or should be) viewed as a Taylor expansion in a small parameter $1-a$, which leads to inflated errors relative to other parameterizations \cite{Colgain:2021pmf}. Marginalizing over higher order terms also leads to inflated errors in any $w_0 w_a$CDM model \cite{Nesseris:2025lke}. Moreover, see \cite{RoyChoudhury:2024wri, Steinhardt:2025znn} for studies highlighting uncertainty in the DESI dynamical DE claims in the extended CPL parameter space.

Even when one restricts attention to independent datasets, a third problem arises: if there is a dynamical DE signal in DESI BAO \cite{DESI:2024mwx, DESI:2025zgx}, when  combined with DESI full-shape (FS) modeling \cite{DESI:2024hhd}, there may be no dynamical DE signal. To appreciate this, note that Fig. 16 of \cite{DESI:2024jis} returns consistent \textit{constant} $\Omega_m$ constraints with the $\Lambda$CDM model within $0.7 \sigma$. Thus, it is imperative to identify BAO data points that are driving one away from the constant $\Omega_m$ behaviour characteristic of the  $\Lambda$CDM model \cite{Colgain:2024mtg}. Note, in DESI DR2 BAO+CMB, one has a dynamical DE signal at $3.1 \sigma$ \cite{DESI:2025zgx}. This deviation is expected to be driven by two effects: i) departures from $\Lambda$CDM behaviour in BAO and ii) a discrepancy in the $\Lambda$CDM parameter $\Omega_m$ between BAO and CMB. As we shall show, DR1 BAO, DR2 BAO and DR2 BAO+CMB fail to confirm late-time accelerated expansion today.

In \cite{Colgain:2024xqj} the consistency of BAO data was studied by translating $D_M(z_i)/r_d$ and $D_H(z_i)/r_d$ constraints into direct constraints on the $\Lambda$CDM parameter $\Omega_m$ at redshift $z_i$. Doing so, it was observed that luminous red galaxy (LRG) data at $z = 0.51$ (LRG1) resulted in {unexpectedly large} $\Omega_m$ values relative to Planck \cite{Planck:2018vyg} at $2.1 \sigma$, whereas LRG data at $z = 0.706$ (LRG2) led to lower $\Omega_m$ values relative to Planck at $1.1 \sigma$. In particular, it was easy to argue that LRG1 data was driving the $w_0 > -1$ signal \cite{DESI:2024mwx, Colgain:2024xqj}. Separately, it was noted that LRG2 distances disagreed with earlier SDSS results \cite{DESI:2024mwx}. This allowed one to argue that statistical fluctuations were at work in LRG1 and LRG2 BAO data \cite{DESI:2024mwx, Colgain:2024xqj} (see also \cite{Dinda:2024kjf, Wang:2024rjd, Wang:2024pui, Chudaykin:2024gol, Liu:2024gfy, Vilardi:2024cwq, Sapone:2024ltl}). These statistical fluctuations disappear when BAO is combined with FS modeling \cite{DESI:2024hhd, DESI:2024jis, Colgain:2024mtg}. 

Given the obvious conflict with Hubble tension and the risk of statistical fluctuations in LRG BAO data when compared to FS modeling, we revisit earlier analysis. As we show, LRG1 continues to return a $\Lambda$CDM $\Omega_m$ value larger than Planck \cite{Planck:2018vyg}, $\Omega_m = 0.315 \pm 0.07$, but at $1.6 \sigma$ removed, it is less anomalous. Moreover, LRG2 BAO has flipped from a lower $\Omega_m$ value to a higher $\Omega_m$ value relative to Planck. This means  that while LRG1 drove the $w_0 > -1$ result in DR1 BAO \cite{Colgain:2024xqj}, the $w_0 > -1$ in DR2 BAO is now driven by both LRG1 and LRG2, but primarily LRG2. On the other hand, whereas the full DESI DR1 BAO dataset preferred a lower $\Omega_m$ value relative to Planck driven by LRGs and emission line galaxies (ELGs), in DESI DR2 BAO the lower $\Omega_m$ value is driven exclusively by the ELGs, in particular ELG1. See \cite{Ormondroyd:2025iaf} for a recent comparison of DESI DR1 and DR2 BAO. {We note that despite the failure to confirm $q_0 < 0$ (without SNe) and the risk of fluctuations, physical modeling of DR2 BAO is underway \cite{Brandenberger:2025hof, Luu:2025fgw, Nakagawa:2025ejs, Paliathanasis:2025dcr, Silva:2025hxw, Wang:2025ljj, Chaussidon:2025npr, Kessler:2025kju, You:2025uon, Pan:2025qwy, Shlivko:2025fgv}.}  Here, we are sympathetic, since the long-standing cosmological constant problem \cite{Weinberg:1988cp} may be alleviated by dynamical DE, e. g. \cite{Yin:2021uus, Yin:2024hba}.       

\section{Analysis}

\begin{figure}[htb]
   \centering
\includegraphics[width=70mm]{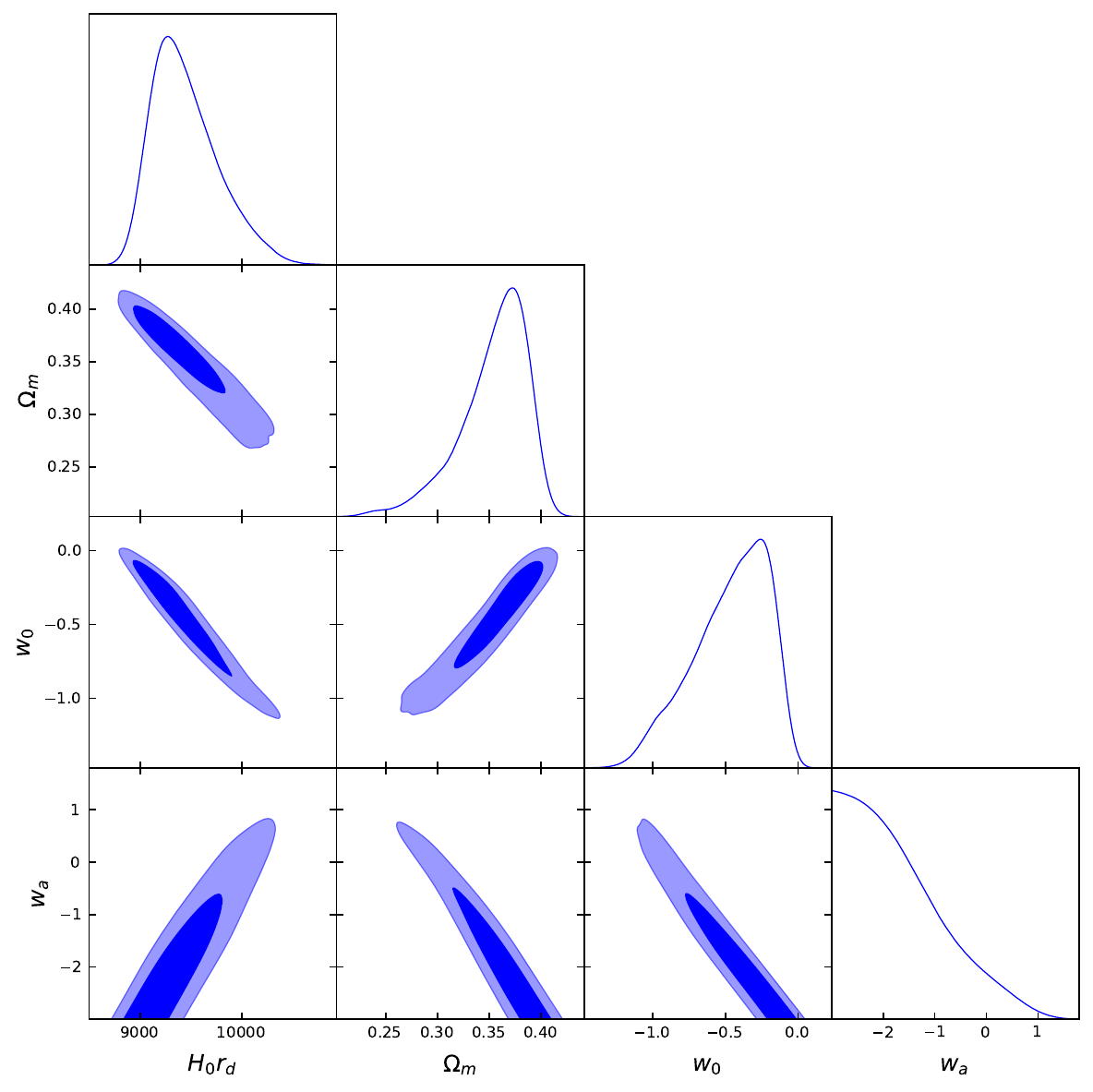}
\caption{CPL model posteriors for DR2 BAO data subject to the DESI priors $w_0 \in [ -3, 1]$, $w_a \in [-3, 2]$ and $w_0 + w_a < 0$. The skewness in $(H_0 r_d, \Omega_m, w_0)$ posteriors comes from the bound $w_a \geq -3$.}
\label{fig:desi_priors} 
\end{figure}

\begin{figure}
   \centering
\includegraphics[width=70mm]{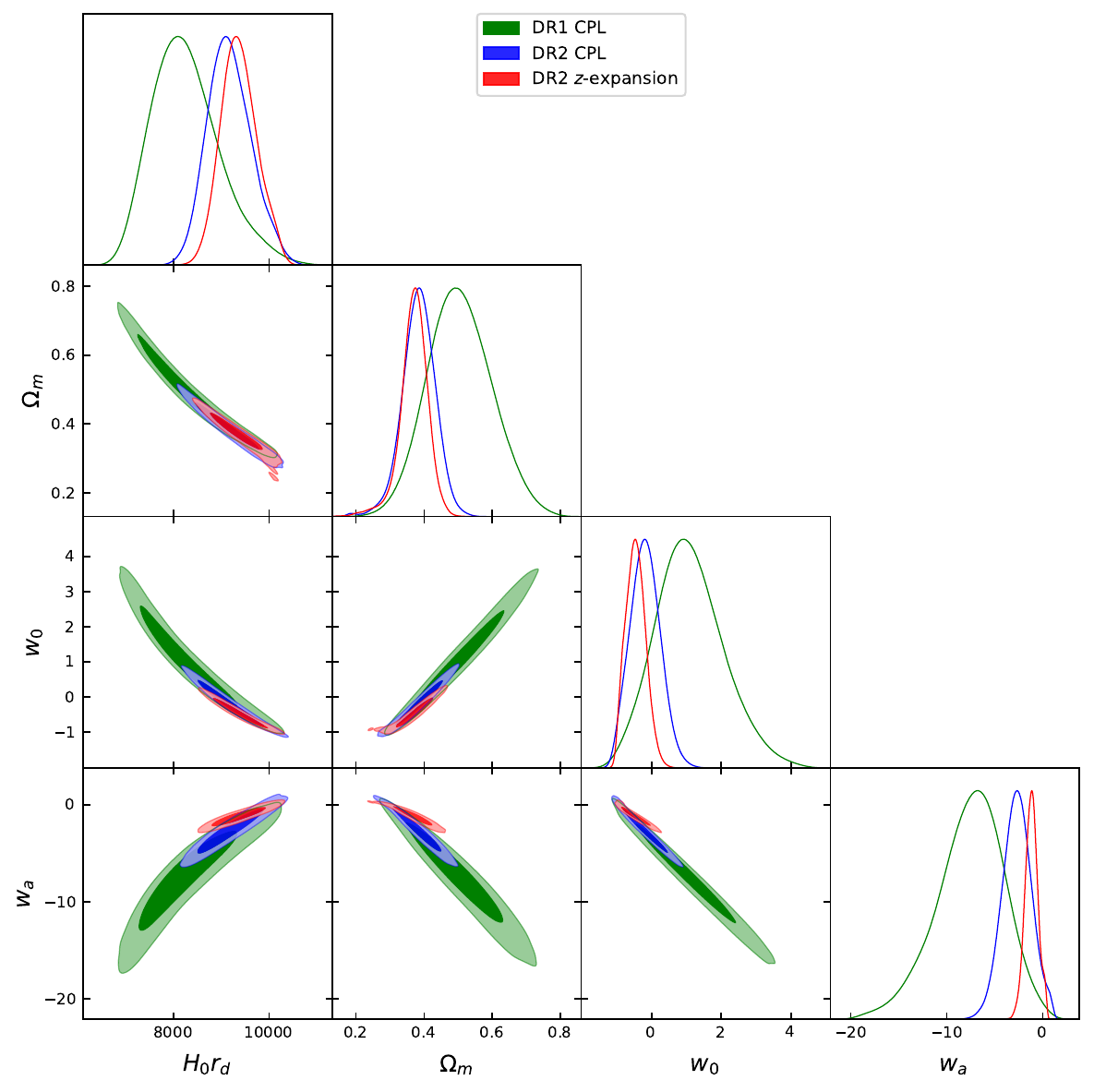}
\caption{Posteriors for $w_0 w_a$CDM models and DESI BAO data subject to the priors $w_0 \in [ -10, 5]$, $w_a \in [-20, 10]$ and $w_0 + w_a < 0$.}
\label{fig:dr1_dr2_w0} 
\end{figure}

We begin with a comment on DESI priors $w_0 \in [-3, 1], w_a \in [-3, 2]$ \cite{DESI:2024mwx, DESI:2025zgx} when DR2 BAO is confronted with the CPL model.\footnote{We do not consider curvature, since anomalies \cite{Handley:2019tkm, DiValentino:2019qzk, vanPutten:2024xwe} have been largely ameliorated by a new Planck likelihood \cite{Tristram:2023haj}.} Note, the choice of priors here is arbitrary as there is no theoretical guidance. In contrast, the DES collaboration use more agnostic priors $w_0 \in [-10, 5], w_a \in [-20, 10]$ \cite{DES:2024jxu}, despite also working with a single observable, namely SNe instead of BAO. By comparing the blue contours in Fig. \ref{fig:desi_priors} (DESI priors) to the blue contours in Fig. \ref{fig:dr1_dr2_w0} (DES priors) one sees that the posteriors are skewed with narrower priors. In particular, it is worth noting that the $H_0 r_d$ posterior is positively skewed, while $\Omega_m$ and $w_0$ posteriors are negatively skewed in Fig. \ref{fig:desi_priors}. When one defines credible intervals using the most common choices, namely i) equal-tailed intervals and ii) highest density intervals with a mode central value, this skewness manifests itself in terms of larger errors coinciding with the longer tails in the posterior. In contrast, DESI appears to quote the mean as a central value \cite{DESI:2025zgx}, $\Omega_m = 0.352^{+0.041}_{-0.018}$, $w_0 = -0.48^{+0.35}_{-0.17}$, so the smaller errors coincide with the longer tails. Replacing the mean value with the mode, we find $\Omega_m = 0.375^{+0.020}_{-0.037}$ and $w_0 = -0.21^{+0.08}_{-0.44}$, but otherwise we agree with the DESI $68 \%$ credible intervals up to small numbers.    

\begin{figure}
   \centering
\includegraphics[width=65mm]{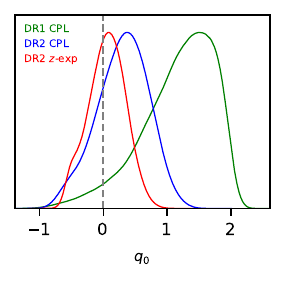}
\caption{$q_0$ posteriors for the CPL posteriors in Fig. \ref{fig:dr1_dr2_w0}.}
\label{fig:q0_models} 
\end{figure}

\begin{table}[htb]
    \centering
    \begin{tabular}{|c|c|c|c|}
    \hline
    \rule{0pt}{3ex} Model+Data & $\Omega_m$ & $w_0$ & $w_a$ \\
    \hline\hline
    \rule{0pt}{3ex} CPL DR1 BAO & $0.502^{+0.098}_{-0.090}$ & $1.03^{+1.0}_{-0.91}$ & $-7.3^{+3.2}_{-3.7}$ \\
    \rule{0pt}{3ex} CPL DR2 BAO & $0.385^{+0.046}_{-0.047}$ & $-0.19^{+0.44}_{-0.43}$ & $-2.7^{+1.5}_{-1.5}$ \\
    \rule{0pt}{3ex} $z$-exp DR2 BAO & $0.372^{+0.036}_{-0.039}$ & $-0.47^{+0.29}_{-0.29}$ & $-1.19^{+0.66}_{-0.67}$ \\
    \hline
    \end{tabular}
    \caption{Posteriors for $w_0 w_a$CDM models and DESI BAO data subject to the priors $w_0 \in [ -10, 5]$, $w_a \in [-20, 10]$ and $w_0 + w_a < 0$.}
    \label{tab:DR1_DR2_CPL_zexp}
\end{table}
In Fig. \ref{fig:dr1_dr2_w0}, having relaxed the priors, we note that the posteriors are visibly more symmetric and Gaussian. Table \ref{tab:DR1_DR2_CPL_zexp} shows the corresponding parameter constraints. Despite $(w_0, w_a)$ values being removed from $\Lambda$CDM expectations, assuming the Planck value $r_d = 147.09$ Mpc to extract $H_0$, we find an age of the Universe $t_{U}$ for CPL fitted to DR1 and DR2 of $t_U = 13.66 \pm 0.08$ Gyrs and $t_U = 13.69 \pm 0.05$ Gyrs, repsectively. Both these values are consistent with a recent estimate from globular clusters, $t_U = 13.57^{+0.28}_{-0.27}$ Gyrs \cite{Valcin:2025luu}. The key point is that relaxing the $(w_0, w_a)$ priors does not lead to a problem for the age of the Universe. Given the symmetric posteriors, it makes little difference how one defines $68\%$ credible intervals, so we opt for equal-tailed intervals based on percentiles. {As Fig. \ref{fig:q0_models} shows}, DR2 BAO data confronted to the CPL model is more consistent than DR1 BAO with late-time accelerated expansion today, which requires 
\begin{equation}
    q_0=\frac12\left[1+3w_0(1-\Omega_m)\right] <0. 
\label{eq:q0}
\end{equation} 
{It is worth noting that $w_a$ does not appear in this expression.} We examine this requirement\footnote{Consistency with the 2011 Physics Nobel Prize, at least at redshift $z=0$, implies a stronger requirement: $q_0 < 0$ at a few $\sigma$ (ideally  $5 \sigma$ or more).  DESI DR2 BAO on its own is not yet of sufficient constraining power to fulfill this requirement within the context of the CPL model. As another comment, our results here can be contrasted with other studies, e.g.  \cite{Gomez-Valent:2018gvm}, where also assuming the CPL model but different data, $q_0 < 0$ is established at $2.5 \sigma$.} by evaluating the expression for $q_0$ on the MCMC chains and check if  $q_0<0$. See Fig. \ref{fig:q0_models} for the $q_0$ posteriors. Converting the MCMC chains into constraints on $q_0$, we find that DR1 BAO confronted to the CPL model rules out $q_0 < 0$ at $95.7 \%$ confidence level, corresponding to $1.7 \sigma$ for a one-sided Gaussian. For DR2 BAO fitted to the CPL model, we find $q_0 < 0$ is ruled out at $76.9 \%$ confidence level ($0.7 \sigma$ for a one-sided Gaussian). See \cite{Wang:2024rjd} for an earlier observation of this tension in DR1. With the improvement in data quality between DR1 and DR2, restrictive priors are no longer required and DR2 BAO shows progress in that $q_0$ is less positive. 

Combining BAO with external datasets further alleviates this problem. However, even for CMB+DR2 BAO, one can convert $\Omega_m =  0.353 \pm 0.021$, $w_0 = -0.42 \pm 0.21$ \cite{DESI:2025zgx} into $q_0 = 0.09 \pm 0.20$, so this result also fails to confirm late-time accelerated expansion today. Our analysis here assumes the $\Omega_m$ and $w_0$ parameters are uncorrelated normally distributed random variables, which can only overestimate the errors on $q_0$, so the discrepancy with $q_0 < 0$ is underestimated. Only when one combines DR2 BAO with SNe can one confirm late-time accelerated expansion today in a meaningful way. Note, for canonical values of $\Omega_m \sim 0.3$, we see that larger values of $w_0$ are problematic from equation (\ref{eq:q0}). In contrast, combining BAO with SNe lowers $w_0$ to avoid any contradiction. 

Interestingly, the same observation can be made for DES SNe \cite{DES:2024jxu}, where $\Omega_m = 0.495^{+0.033}_{-0.043}$, $w_0 = -0.36^{+0.36}_{-0.30}$ can be converted into $q_0 = 0.23 \pm 0.27$. \footnote{Whether one symmetrizes the errors by adopting the largest error as the standard deviation for a normal distribution or models the aysmmetry through distinct normal distributions for values larger and smaller than the central value, there is no appreciable difference to the outcome.} Given that DESI BAO and DES SNe have a higher effective redshift than traditional SNe samples, this may highlight the difficulty constraining $q_0$ to the widely accepted value, $q_0 < 0$, when the datasets are poorly anchored at lower redshifts. However, theoretically, there is a more sensational possibility. The Trans-Planckian Censorship
Conjecture (TCC) \cite{Bedroya:2019snp, Bedroya:2019tba} implies that late-time accelerated expansion must come to an end \cite{Brandenberger:2019jbs, Brandenberger:2025hof, Li:2025cxn}, so one could interpret $q_0 > 0$ as observational support. 

As an added check, we replace the CPL model with the analogous $z$-expansion model, $w_{\textrm{DE}}(z) = w_0 + z \, w_a$, which may be viewed as a Taylor expansion in $z$ instead of $1-a = z/(1+z)$ for CPL. One finds  a central value more consistent with late-time accelerated expansion today while all the errors shrink accordingly. This reduction in the errors is expected, since as explained in \cite{Colgain:2021pmf}, $z$ is a larger expansion parameter than $1-a$ and this allows the data to place stronger constraints on $w_a$, which in turn better constrains the remaining parameters. Objectively, the CPL model is a dynamical DE model that is paradoxically less sensitive at lower redshifts. Changing the model from CPL to the $z$-expansion model, we find that $q_0 < 0$ is ruled out at $57.6 \%$ confidence level ($0.2 \sigma$ for a one-sided Gaussian). See Fig. \ref{fig:q0_models} for the posterior. 

\begin{figure}
   \centering
\includegraphics[width=80mm]{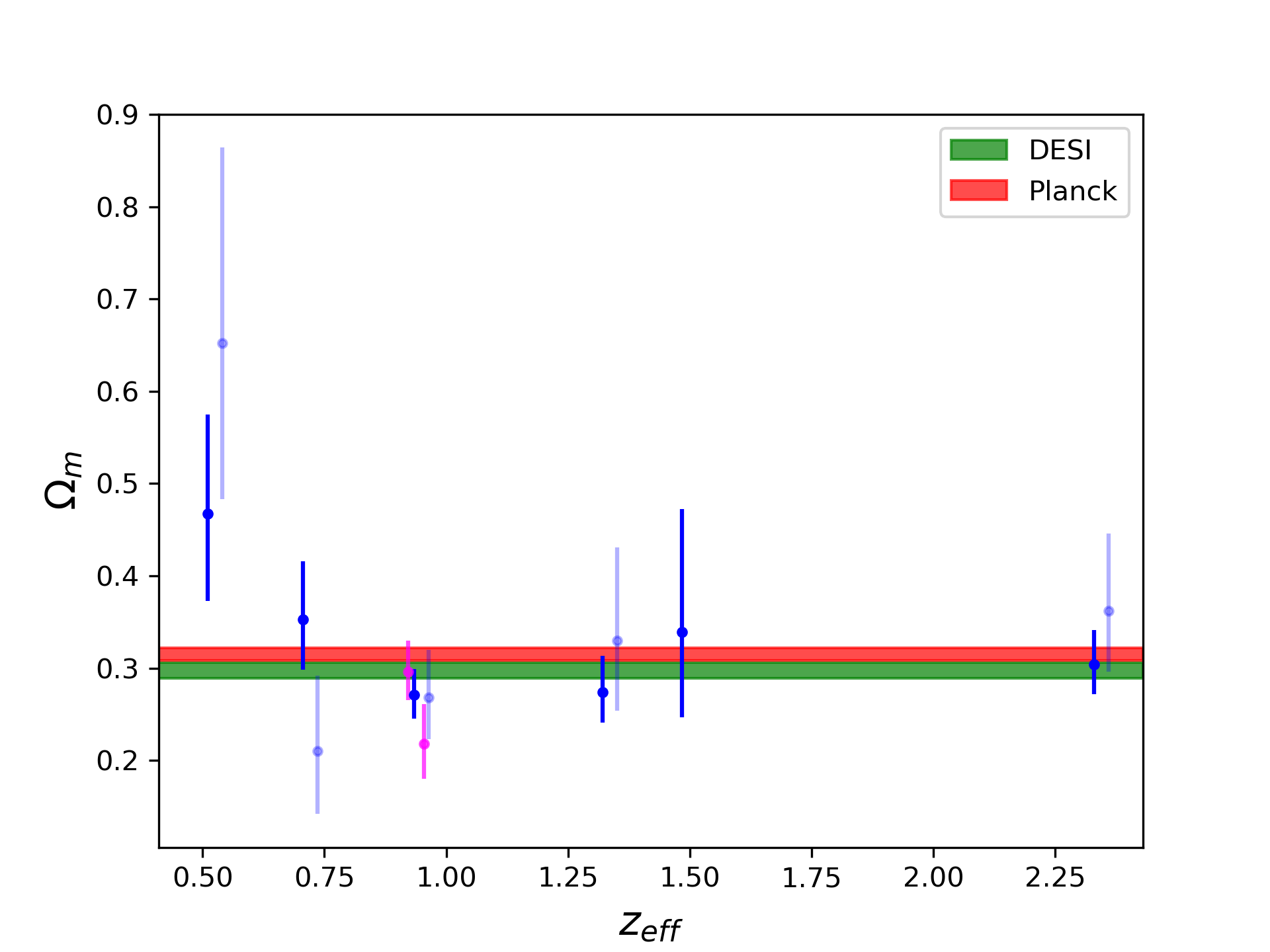}
\caption{Differences in the $\Lambda$CDM $\Omega_m$ constraints from individual tracers between DR1 in faded blue and DR2 in blue. The red and green bands denote Planck and full DESI DR2 BAO sample constraints on $\Omega_m$. In magenta we separate LRG3 and ELG1 constraints. We have displaced redshifts for visually purposes.}
\label{fig:DR1vDR2} 
\end{figure}

Our next task is to identify which data points in DESI DR2 BAO  are driving the $w_0> -1$ result. One could alternatively focus on the complementary parameter $w_a$. The logic we use is that deviations from $\Lambda$CDM captured by the $w_0 w_a$CDM model must be evident as differences in $\Omega_m$ at different redshifts in the flat $\Lambda$CDM model. We follow the methodology of \cite{Colgain:2024xqj}, where for each effective redshift in Table IV of \cite{DESI:2025zgx} with both $D_{M}(z_i)/r_d$ and $D_H(z_i)/r_d$ constraints, we construct a $2 \times 2$ covariance matrix with the correlation $r$, generate 10,000 $(D_{M}(z_i)/r_d, D_H(z_i)/r_d)$ pairs, and solve the following equation for $\Omega_m$ for each pair:
\begin{equation}
\label{eq:DMDH}
    \frac{D_M(z)/r_d}{D_H(z)/r_d} = E(z) \int_0^z \frac{1}{E(z^{\prime})} \dd z^{\prime}. 
\end{equation}
This ratio only depends on $\Omega_m$ in the $\Lambda$CDM model where $E(z) = \sqrt{1-\Omega_m + \Omega_m (1+z)^3}$. It has been checked that this methodology leads to comparable errors to Markov Chain Monte Carlo (MCMC) \cite{Colgain:2024xqj}. The result of this exercise is shown in Table \ref{tab:DESI_OM_DR2} and Fig. \ref{fig:DR1vDR2}, where only the lower redshift bright galaxy sample (BGS) is omitted. For each entry in Table \ref{tab:DESI_OM_DR2} we confirmed that MCMC gives comparable results, as expected. As remarked in  \cite{DESI:2025zgx}, omitting BGS is not a problem as we do not expect strong constraints on $\Omega_m$ at lower redshifts. We include the results of LRG3 and ELG1 for completeness in Table \ref{tab:DESI_OM_DR2}, but observe that they are not independent from the LRG3+ELG1 entry \cite{DESI:2025zgx}.

\begin{table}
    \centering
    \begin{tabular}{|c|c|c|}\hline
    \rule{0pt}{3ex} tracer & $z_{\textrm{eff}}$ & $\Omega_m$ \\
    \hline\hline
    \rule{0pt}{3ex} LRG1 & $0.510$ & $0.467^{+0.11}_{-0.094}$ \\
    \rule{0pt}{3ex} LRG2 & $0.706$ & $0.353^{+0.063}_{-0.055}$ \\
    \rule{0pt}{3ex} LRG3+ELG1 & $0.934$ & $0.271^{+0.028}_{-0.026}$ \\
    \rule{0pt}{3ex} ELG2 & $1.321$ & $0.274^{+0.039}_{-0.033}$ \\
    \rule{0pt}{3ex} QSO & $1.484$ & $0.339^{+0.133}_{-0.092}$ \\
    \rule{0pt}{3ex} Lyman-$\alpha$ QSO & $2.330$ & $0.304^{+0.037}_{-0.032}$ \\
    \hline 
    \rule{0pt}{3ex} LRG3 & $0.922$ & $0.296^{+0.034}_{-0.031}$ \\
    \rule{0pt}{3ex} ELG1 & $0.955$ & $0.218^{+0.043}_{-0.038}$ \\
    \hline
    \end{tabular}
    \caption{Constraints on the $\Lambda$CDM parameter $\Omega_m$ from individual tracers.}
    \label{tab:DESI_OM_DR2}
\end{table}

As remarked earlier, LRG1 data now leads to an $\Omega_m$ value that is more consistent with the traditional $\Omega_m \sim 0.3$. Relative to the Planck $\Omega_m$ value in red in Fig. \ref{fig:DR1vDR2}, we see that all constraints intersect the red strip except for LRG1 and LRG3+ELG1. Splitting the LRG3+ELG1 constraint into its components, one sees that this is due to the low $\Omega_m$ value ELG1 tracer. Shifting the red horizontal strip downwards to the location of the green strip corresponding to the DESI DR2 BAO $\Omega_m$ constraint for the full sample, one can see that this provides visually a better fit to all constraints, except the LRG1 constraint and ELG1 constraint once separated from LRG3. LRG1 returns a larger $\Omega_m$ value at $1.8 \sigma$ and ELG1 returns a smaller $\Omega_m$ value also at $ 1.8 \sigma$.\footnote{More carefully, removing the LRG1 and ELG1 data from the remaining data we have $\Omega_m = 0.293 \pm 0.009$ and $\Omega_m = 0.301 \pm 0.009$, respectively, so the discrepancies with the LRG1 and ELG1 constraints in Table \ref{tab:DESI_OM_DR2} and the remaining data are $1.8 \sigma$ and $1.9 \sigma$.} This makes LRG1 the most prominent outlier once again, but ELG1, when separated from LRG3, is on par. It should be noted that in contrast to DR1, where LRG2 data was contributing to the lower $\Omega_m$ in the full BAO sample, we can now confirm that this is driven exclusively by ELG data. Thus, if one wanted to restore consistency with Planck-$\Lambda$CDM, it would be enough to explore ELG tracers, in particular ELG1, for potential systematics. That being said, while DR2 BAO prefers an $\Omega_m$ value $1.6 \sigma$ lower than Planck, removing ELG1 leads to an $\Omega_m$ value that is $1.2 \sigma$ lower. This can be compared with DR1 BAO, which led to an $\Omega_m$ value $1.2 \sigma$ lower than Planck. It is also worth noting that the LRG3+ELG1 constraint has exhibited the smallest shift in $\Omega_m$ value between DR1 and DR2.   

We now ask again the question posed in \cite{Colgain:2024xqj}: is it possible to remove a single constraint, for example LRG1, and recover $w_0 = -1$ (or $w_a = 0$) within $1 \sigma$? In Table \ref{tab:noLRG_ELG} we document the effect of removing LRG1, LRG2 and ELG1 data, which are the most obvious outliers that could contribute to the $w_0 > -1$ result in the full sample. From the CPL DR2 BAO entry in Table \ref{tab:DR1_DR2_CPL_zexp} we observe that $w_0 = -1$ is  $1.9 \sigma$ from the $w_0$ value preferred by the full sample. From Table \ref{tab:noLRG_ELG}, removing LRG1, ELG1 and LRG2 in turn, we find $w_0 =-1$ is $1.9 \sigma$, $1.7 \sigma$ and $1.3 \sigma$ removed, respectively.

Interestingly, we find that the removal of LRG1 pushes one further into a regime where there is no accelerated expansion today, but has no effect on the preference for $w_0 > -1$. That being said, {once LRG1 is removed we find that $q_0 < 0$ is disfavoured at $87.3 \%$ confidence level ($1.1 \sigma$ for a one-sided Gaussian) compared to $76.9\%$ for the full sample}. Removing ELG1 brings one marginally closer to $w_0 = -1$, confirming that this outlier has some influence on the $w_0 > -1$ result. In contrast, we find that removing LRG2 brings $w_0$ closest to $w_0 = -1$. What this implies is that the $w_0 > -1$ deviation from $\Lambda$CDM in DR2 BAO is driven mainly by LRG2 data, whereas in DR1 BAO, this was driven by LRG1 data. 

\begin{figure}
   \centering
\includegraphics[width=80mm]{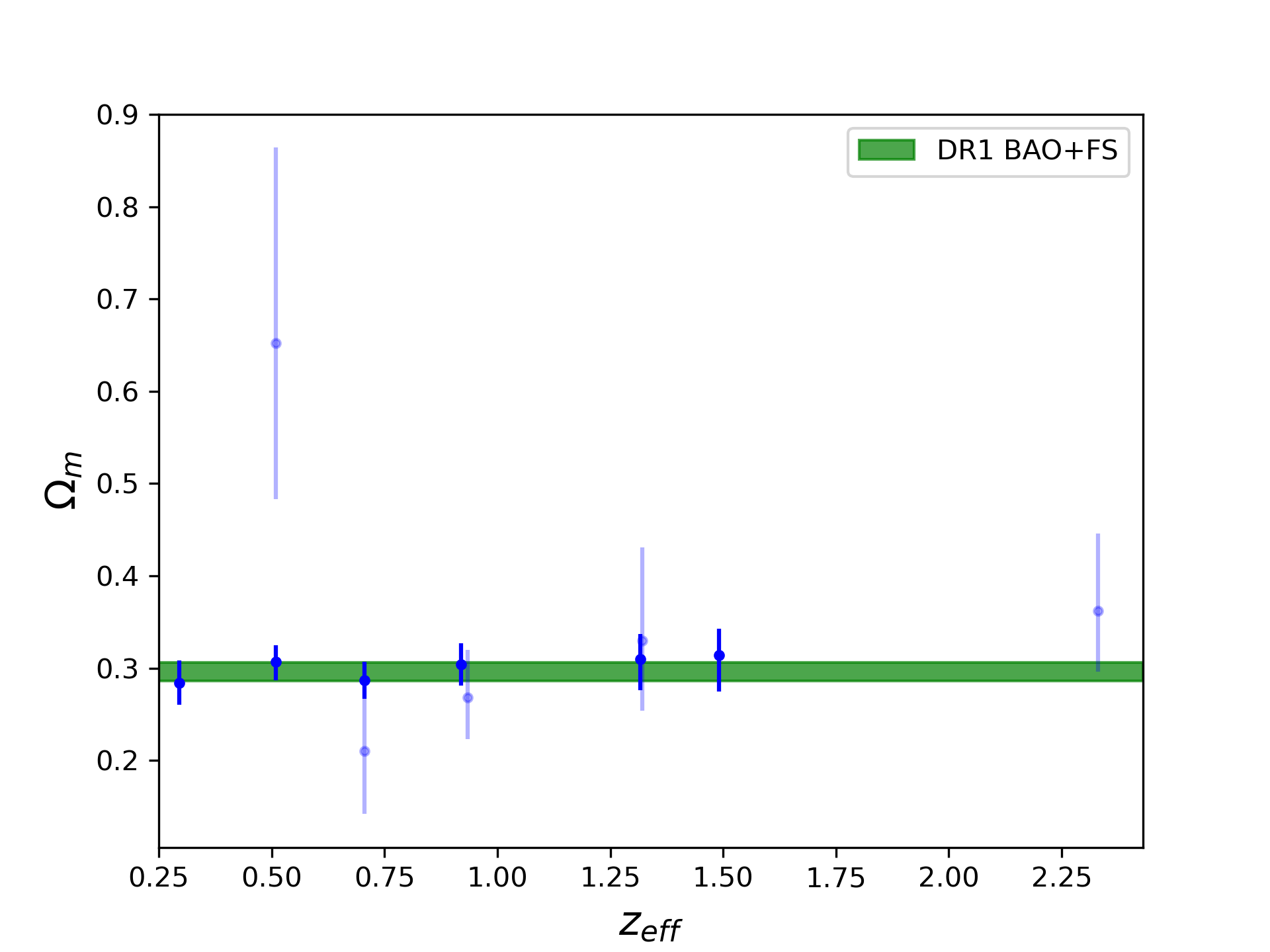}
\caption{DESI DR1 BAO constraints on the $\Lambda$CDM parameter $\Omega_m$ in faded blue relative to the DESI DR1 BAO+FS modeling constraints in blue. The green strip denotes the constraint on the full sample from BAO+FS and it can be confirmed that all constraints show excellent agreement. Ideally, BAO and FS modeling should agree on $\Omega_m$.}
\label{fig:BAOvsFS} 
\end{figure}

In summary, our analysis exposes a $q_0$ sign problem and evident fluctuations in DESI DR2 BAO. As is clear from Fig. \ref{fig:DR1vDR2}, LRG3+ELG1 shows excellent agreement between DR1 and DR2, while ELG2 and Lyman-$\alpha$ QSO also show good agreement. On the other hand, LRG1 has become more consistent with Planck. LRG1 and ELG1 are outliers at $1.8 \sigma$ with respect to the value of $\Omega_m$ preferred by the full DR2 BAO sample. However, the most interesting difference between DR1 and DR2 is the shift in LRG2, which means it goes from contributing to the lower $\Omega_m$ preferred by the full sample relative to Planck, to the tracer that is most sharply driving the dynamical DE signal (see Fig. \ref{fig:DR1vDR2}). In Fig. \ref{fig:BAOvsFS} we remind the reader that despite the potential for statistical fluctuations in DESI BAO data, when DESI DR1 BAO is combined with FS modeling, there are no obvious outliers and all constraints intersect the $\Omega_m$ value for the full BAO+FS dataset \cite{DESI:2024hhd, DESI:2024jis}, i. e. the equivalent of LRG1 and ELG1 BAO constraints do not exist. As a result, there is no signal of dynamical DE in DESI DR1 data alone. That being said, DESI DR1 BAO+FS data prefers a value for the $\Lambda$CDM parameter $\Omega_m$ that is $1.6 \sigma$ lower than the Planck value. Admittedly, this constant shift in $\Omega_m$ challenges concordance, but claims that $\Omega_m$ is not a constant in $\Lambda$CDM cosmology, and is in fact redshift dependent, go back to 2022 \cite{Colgain:2022nlb, Colgain:2022rxy}. {The main point here is that a non-constant $\Omega_m$ $\Lambda$CDM parameter, while pointing to model breakdown, does not immediately imply a dynamical DE sector.}

\begin{table}[htb]
    \centering
    \begin{tabular}{|c|c|c|c|}\hline
    \rule{0pt}{3ex} Data & $\Omega_m$ & $w_0$ & $w_a$ \\
    \hline\hline
    \rule{0pt}{3ex} no LRG1 & $0.418^{+0.066}_{-0.062}$ & $0.20^{+0.68}_{-0.63}$ & $-4.1^{+2.1}_{-2.3}$ \\
    \rule{0pt}{3ex} no ELG1 & $0.381^{+0.045}_{-0.049}$ & $-0.27 \pm 0.43$ & $-2.4 \pm 1.5$ \\
    \rule{0pt}{3ex} no LRG2 & $0.363^{+0.049}_{-0.054}$ & $-0.42^{+0.47}_{-0.46}$ & $-2.0^{+1.7}_{-1.6}$ \\
    \hline
    \end{tabular}
    \caption{Constraints on the CPL model from the full sample with LRG1, ELG1 and LRG2 data removed.}
    \label{tab:noLRG_ELG}
\end{table}

\section{Discussion}
Any dynamical DE signal with $w_{0} > -1$ cannot be the final word on a replacement for the $\Lambda$CDM model because the $w_0-H_0$ anti-correlation is problematic in the face of $H_0 > 70$ km/s/Mpc determinations \cite{Vagnozzi:2018jhn, Vagnozzi:2019ezj, Alestas:2020mvb, Banerjee:2020xcn, Lee:2022cyh}. A key point is that larger than expected local $H_0$ values are observed across multiple observables. As a result, one infers that there must be something wrong with the DESI dynamical DE claim \cite{DESI:2024mwx, DESI:2024hhd, DESI:2025zgx}, at least in its current form. It is true that the CPL model fits DESI+CMB+SNe and DESI+CMB datasets better than $\Lambda$CDM in a statistically significant manner, but if it contradicts Hubble tension, {or worse, fails to confirm late-time accelerated expansion today}, one should rethink.\footnote{{Physics traditionally makes progress by identifying and resolving contradictions.}} That being said, it should be borne in mind that Hubble tension is an expected harbinger of new physics beyond $\Lambda$CDM. Constructing a science case against $\Lambda$CDM when it is confronted to diverse observables is relatively easy \cite{Chaudhary:2025uzr}.

In this letter we looked at the improvements in DESI BAO data between DR1 \cite{DESI:2024mwx} and DR2 \cite{DESI:2025zgx}. When the data is confronted to the CPL model, we noted that the narrow $w_a \in [-3, 2]$ priors employed by the DESI collaboration cut off the $w_a$ posterior, giving rise to skewed posteriors that become more symmetric once the priors are relaxed. In agreement with earlier work \cite{Wang:2024rjd}, with relaxed priors we confirmed that DESI DR1 BAO is inconsistent with late-time accelerated expansion today ($q_0 < 0$). In the upgrade to DESI DR2 BAO, we note that $q_0 < 0$ cannot be confirmed, even when one combines DR2 BAO with CMB. Thus, we have a $3.1 \sigma$ deviation from $\Lambda$CDM \cite{DESI:2025zgx},\footnote{Note, DESI DR2 BAO and CMB have an approximate $1.6 \sigma$ discrepancy in the $\Lambda$CDM parameters $\Omega_m$. This can only translate into a $3.1 \sigma$ deviation from $\Lambda$CDM in the CPL model \cite{DESI:2025zgx} if BAO data has an inherent dynamical DE signal. Replacing BAO with FS modeling, where there is no inherent dynamical DE signal, the statistical significance of the deviation must drop.} yet cannot confirm $q_0 < 0$. This may have tantalising implications for theoretical speculation \cite{Brandenberger:2019jbs, Brandenberger:2025hof, Li:2025cxn}, but more mundanely may simply highlight a problem fitting the CPL model to datasets with a high effective redshift. In particular, it is plausible that a statistical fluctuation at the higher redshifts probed by LRG, which is otherwise relatively benign, explains this unexpected result in the extrapolation to $z=0$.

Given that DESI DR1 BAO data are prone to fluctuations \cite{DESI:2024mwx, Colgain:2024xqj}, and the dynamical DE signal may have hinged on an isolated LRG1 tracer \cite{Colgain:2024xqj}, we revisited earlier analysis that translates $(D_M(z_i)/r_d, D_H(z_i)/r_d)$ pairs at redshift $z_i$ into constraints on the $\Lambda$CDM parameter $\Omega_m$. {This is an important exercise as the statistically significant deviation from $\Lambda$CDM reported in DR2 BAO+CMB at $3.1 \sigma$ \cite{DESI:2025zgx} {but $q_0 > 0$} must be due to BAO data.} 

We observe that LRG1, which was the most prominent outlier in DR1 BAO \cite{Colgain:2024xqj}, is now consistent with a lower $\Omega_m$ value in the $\Lambda$CDM model. It is still jointly the most prominent outlier at $1.8 \sigma$ removed from the rest of the dataset. {In addition, we see that ELG1 prefers an $\Omega_m$ value $1.8 \sigma$ lower than the full sample.} Moreover, we find that LRG2 returns an $\Omega_m$ value larger than Planck in DR2 in contrast to the smaller value in DR1. We now confirm that ELG data is solely responsible for the lower $\Omega_m$ relative to Planck in the full DR2 BAO sample. Finally, we show that LRG2 and neither LRG1 nor ELG1 is now most responsible for the $w_0 > -1$ dynamical DE signal. Ref. \cite{Goldstein:2025epp} also arrives at the same conclusion through different methods. See also \cite{Chaudhary:2025pcc} with similar conclusions but a stronger overlap in methods.

Evidently, fluctuations still persist in DESI BAO data and we have yet to see convergence in constraints for both LRG1 and LRG2, although high redshift tracers show good to excellent agreement between DR1 and DR2. {In addition, ELG1 results in an unusually low $\Omega_m$ value.} It will be interesting to see if any fluctuations remain when DR2 BAO is combined with DR2 FS modeling, as there is no dynamical DE signal when DR1 BAO is combined with DR1 FS modeling \cite{DESI:2024jis}. 

{It is important to note that a dynamical DE signal can have two origins. In BAO+CMB+SNe combinations, differences in the $\Lambda$CDM parameter $\Omega_m$ between datasets at different effective redshifts can manifest as a dynamical DE signal even if there is no dynamical DE signal in BAO and SNe independently. See \cite{Colgain:2024mtg} for consistency checks of this possibility. See also \cite{Gialamas:2024lyw, Huang:2024qno, Efstathiou:2024xcq, Huang:2025som} for speculation that a mismatch between DESI and SNe at lower redshifts is responsible for the signal. The second possibility is that there is a genuine dynamical DE signal in these independent datasets, one currently seen in DES SNe \cite{DES:2024jxu}, DESI BAO \cite{DESI:2024mwx, DESI:2025zgx}, but importantly not DESI FS modeling \cite{DESI:2024hhd, DESI:2024jis}. The community needs to separate these two possibilities and study them independently. Nevertheless, no matter how one looks at it, $w_0 > -1$ has a Hubble tension problem \cite{Vagnozzi:2018jhn, Vagnozzi:2019ezj, Alestas:2020mvb, Banerjee:2020xcn, Lee:2022cyh}, which risks making any discussion moot.}

\section*{Acknowledgements} We would like to thank Eleonora Di Valentino and Sunny Vagnozzi for correspondence on credible intervals. {We thank Adri\`a G\'omez-Valent for helpful comments on the manuscript.} This article/publication is based upon work from COST Action CA21136 – “Addressing observational tensions in cosmology with systematics and fundamental physics (CosmoVerse)”, supported by COST (European Cooperation in Science and Technology). M.M. ShJ would like to thank Iranian National Science Foundation (INSF) research chair grant no. INSF 4045163. L. Yin was supported by Natural Science Foundation of Shanghai
24ZR1424600.

\appendix

\section{Dynamical DE and $H_0$ tension}

On the assumption of Gaussian errors, we quantify the tension between $H_0$ values inferred from DESI DR2 BAO combined with external datasets and the SH0ES value, $H_0 = 73.04 \pm 1.04$ km/s/Mpc \cite{Riess:2021jrx}. What we want to demonstrate is that the further $w_0>-1$ deviates from $w_0 = -1$, the greater the tension with SH0ES becomes. This is a strong indication that one could never restore concordance with the DESI collaboration's claimed dynamical DE signal, since one can only deviate from flat $\Lambda$CDM by exacerbating a more established tension. In Table \ref{tab:DESI_H0tension} we document the results, where $H_0$ and $w_0$ constraints are reproduced from Table V of \cite{DESI:2025zgx}.

\begin{table*}[htb]
    \centering
    \begin{tabular}{|l|c|c|c|c|}\hline
    \rule{0pt}{3ex} Data & $H_0$ [km/s/Mpc] & $H_0$ tension & $w_0$ & $w_0$ tension \\
    \hline\hline
    \rule{0pt}{3ex} DESI+CMB+Pantheon+  & $67.51 \pm 0.59$ & $4.6 \sigma$ & $-0.838 \pm 0.055$ &  $2.9 \sigma$ \\
    \rule{0pt}{3ex} DESI+CMB+Union3 & $65.91 \pm 0.84$ & $5.3 \sigma$ & $-0.667 \pm 0.088$ & $3.8 \sigma$  \\
    \rule{0pt}{3ex} DESI+CMB+DES & $66.74 \pm 0.56$ & $5.3 \sigma$ & $-0.752 \pm 0.057$ & $4.4 \sigma$\\
    \hline
    \end{tabular}
    \caption{DESI results reproduced from Table V of \cite{DESI:2025zgx} alongside the $H_0$ tension with SH0ES \cite{Riess:2021jrx} and the $w_0$ tension with $\Lambda$CDM ($w_0=-1$). $H_0$ tension monotonically increases with $w_0$ tension.}
    \label{tab:DESI_H0tension}
\end{table*}

First, it is worth noting that the data are constraining enough that both $H_0$ and $w_0$ errors are consistent with Gaussian posteriors. In contrast, as can be seen from Table V of \cite{DESI:2025zgx}, the corresponding $w_a$ errors are not, so we ignore them. Secondly, as the central value for $w_0$ increases from $w_0 = -1$, the central value of $H_0$ decreases, thus the central values are strongly anti-correlated. It remains to factor in the errors to establish a statistical significance. Incorporating the errors, one sees that an increase in the statistical significance of the $w_0$ deviation from $w_0 = -1$ typically corresponds to an increase in $H_0$ tension. There is one exception to this when one switches out Union3 SNe \cite{Rubin:2023ovl} with DES SNe \cite{DES:2024jxu} and the $w_0$ tension increases, but this is not reflected in an increase in $H_0$ tension. Nevertheless, it is true that an increase in $H_0$ tension driven by $w_0$ tension is monotonic, since the former never decreases.

\section{Simple argument on $w(z)$ and $H_0$}

In \cite{Vagnozzi:2018jhn, Vagnozzi:2019ezj, Alestas:2020mvb, Banerjee:2020xcn, Lee:2022cyh} it has been shown pretty generally that increasing $w(z) = -1$ ($\Lambda$CDM) to $w(z) > -1$ (quintessence) lowers $H_0$. Here is an analytic argument. In general, the DE density $X(z)$ is related to the equation of state $w(z)$ through:
\begin{equation}
    X(z) = \exp \left( 3 \int_0^z \frac{1+w(z^{\prime})}{1+z^{\prime}} \dd z^{\prime} \right).  
\end{equation}
Observe that $w(z) > -1 \Rightarrow X(z) > 1$ and for $\Lambda$CDM $X(z) =1$. Consider an  $X(z)$CDM dynamical DE model with Hubble parameter 
\begin{equation}
    H(z)^2 = H_0^2 (1-\Omega_m) X(z) + H_0^2 \Omega_m (1+z)^3, 
\end{equation}
where we restrict our attention to the late Universe. By definition the DE term is not supposed to be relevant at high redshift, as is the case in $\Lambda$CDM, hence the combination $H_0^2 \Omega_m$ is expected to be essentially constrained the same in $\Lambda$CDM and in any replacement $X(z)$CDM model. It is easy to see now that if $H_0^2 \Omega_m$ is a constant, but one increases $X(z)$ through a theoretical choice, then $H_0^2 (1-\Omega_m)$ with constant $H_0^2 \Omega_m$, and hence $H_0$, must decrease to find a fit no worse than $\Lambda$CDM to the data. Note, the $X(z)$CDM model with $X(z) > 1$ can fit the data better than $\Lambda$CDM, but the price one pays is a lower $H_0$. This guarantees that one cannot resolve $H_0$ tension, and errors aside, most likely makes the problem worse.

Our argument here covers analysis in \cite{Banerjee:2020xcn, Lee:2022cyh}. In \cite{Banerjee:2020xcn} observed data was studied, and while quintessence models that raised $H_0$ were found, they all fitted the data worse than $\Lambda$CDM. Despite CMB data not being utilised, one infers that the combination $H_0^2 \Omega_m$ must have remained constant between $\Lambda$CDM and quintessence fits to a combination of BAO, cosmic chronometer and SNe data. In contrast, \cite{Lee:2022cyh} worked with mock DESI forecast data up to higher redshifts $z \leq 3.55$. We emphasize that in both these exercises no exceptions were found; an increase in $w(z)$ always led to a decrease in $H_0$. As can be seen from the analytic argument, this is guaranteed provided both $\Lambda$CDM and the replacement DE model with $w(z) > -1$ lead to the same constraints on $H_0^2 \Omega_m$. The non-constancy of $H_0^2 \Omega_m$ is the only loophole.

\bibliography{refs}

\end{document}